\documentclass[12pt]{article}

\usepackage{amsmath}
\usepackage{amssymb}
\usepackage{amsthm}
\usepackage{mathrsfs}
\usepackage[T1]{fontenc}
\usepackage{enumerate}
\usepackage{color}

\def\eqn{equation}
\def\cond{condition}

\def\tfn{transformation}

\def\sm{sigma model}

\def\dd{Drinfel'd double}

\def\4diml{four-dimensional}
\def\bkg{background}

\def\-1{^{-1}}
\def\half{\frac{1}{2}}
\def\coor{coordinate}

\def\e{{e}}

\def\cd{{\mathfrak d}}
\def\cg{{\mathfrak g}}
\def\tcg{\tilde{\mathfrak g}}
\def\hcg{\hat{\mathfrak g}}
\def\bcg{\bar{\mathfrak g}}

\def\ytoxhat{\Big|_{y=\hd.\hat x}}

\def\wt{\tilde}
\def\wh{\widehat}

\def\sm{sigma model}
\def\PL{Poisson--Lie }
\def\pltp{Poisson--Lie T-pluralit}

\def\sugra{Generalized Supergravity Equation}

\def\cf{{\mathcal {F}}}

\newcommand{\unit}{\mathbf{1}}
\newcommand{\nul}{\mathbf{0}}

\newcommand{\M}{\mathscr{M}}
\newcommand{\D}{\mathscr{D}}
\newcommand{\G}{\mathscr{G}}
\newcommand{\tG}{\widetilde{\mathscr{G}}}
\newcommand{\hG}{\widehat{\mathscr{G}}}
\newcommand{\bG}{\bar{\mathscr{G}}}

\newcommand{\hd}{\hat{d}}

\newcommand{\N}{\mathscr{N}}

\newcommand{\wb}{\bar}

\def\cor{}

\begin{document}
\title{Compatibility of  Poisson--Lie transformations and symmetries of Generalized Supergravity Equations}
\author{Ladislav Hlavat\'y\footnote{hlavaty@fjfi.cvut.cz}
\\ {\em Faculty of Nuclear Sciences and Physical Engineering,}
\\ {\em Czech Technical University in Prague,}
\\ {\em Czech Republic}
}
\maketitle
\abstract{We investigate two types of transformations that keep  NS-NS Generalized Supergravity Equations satisfied : $\chi$-symmetry \eqref{ambig X} that shifts dilaton and gauge transformations \eqref{gauge tfn lambda} that change both dilaton and vector field $J$. Due to these symmetries there is a large set of dilatons and vector fields $J$ that (for a fixed metric and B-field) satisfy  Generalized Supergravity Equations but only some of them can be be used as input for Poisson--Lie transformations. Conditions that define the admissible dilatons are given and examples are presented.
}
\tableofcontents
\section{Introduction}
\PL \tfn\ of solutions of \sugra s must include 
dilatons and Killing fields. 
Formula  for \tfn\ of dilaton field $\Phi$ accompanying \PL \tfn\ \cite{klise,kli:pltd} of sigma model \bkg\ $\cf={\mathcal {G}}+{\mathcal {B}}$ 
was given in \cite{unge:pltp} (see also \cite{dehato2,BorWulff:DFT,Diego}).    Formula for \PL \tfn\ of Killing field $J$ 
was given in \cite{demhast,saka2}. Later it turned out that the the latter formula works well only for non-Abelian T-duality, i.e. for \PL \tfn s of \sm s with isotropic \bkg s,  and it was extended in \cite{hlape:formula for J} for other type of \tfn s. Unfortunately, 
applicability of these formulas is dependent on choice of initial dilatons related by symmetries of solution space of \sugra s. It turns out 
that transformed dilatons and vector fields $J$ 
keep validity of 
\sugra s in dependence on the choice on the initial dilaton. Examples of these cases are given below.

Goal of this paper is to discuss compatibility of 
\PL \tfn\ of dilaton and Killing field with \tfn s that leave invariant (NS-NS part of) \sugra s and give conditions for applicability of the \tfn s.

\section{\PL \tfn s and \\ \sugra s }
\PL\ duality/plurality is based on the possibility to pass between various decompositions of \dd\ $\D$ that generates \bkg\ of investigated sigma models. It is a $2d$-dimensional Lie group whose Lie algebra $\cd$ can be decomposed into double cross sum of Lie subalgebras $\cg$ and $\tcg$ that are maximally isotropic with respect to non-degenerate symmetric bilinear ad-invariant form $\langle.,.\rangle$. \dd\ with so called Manin triple $(\cd, \cg, \tcg)$ and Lie subgroups $\G, \tG$ corresponding to $\cg, \tcg$ is denoted by $(\G|\tG)$. 

Assume that there is $d$-dimensional Lie group $\G$ whose action on $\M$ is smooth, proper and free. The action of $\G$ is transitive on its orbits, hence we may locally consider
$\M\approx (\M/\G) \times \G = \N \times \G$, $\dim \M = \dim \N + \dim \G = n+d$ and introduce adapted
coordinates
\begin{equation}\label{adapted}
 \{y^\mu\}=\{s_\alpha,y^a\},
\end{equation}
where  $y^a$ are group coordinates and $s_\alpha$ label the orbits of $\G$, \begin{equation}
\mu=1, \ldots\dim \M,\ \alpha=1, \ldots,n = \dim \N ,\ \ a=1,
\ldots, d = \dim \G. 
\end{equation}
Coordinates $s_\alpha$ are treated as ''spectators'' as they do not participate in \PL\ \tfn s.

\PL\ dualizable sigma models on $\N \times \G$ are given by tensor field \begin{equation}\label{cf}
\cf=\cf(y^\mu)=\cf(s_\alpha,y^a),
\end{equation}
that satisfy
\cite{klise,kli:pltd}\begin{equation}
 ({\cal L}_{v_i}\cf)_{\mu\nu}= \cf_{\mu\kappa}v^\kappa_j\tilde f_i^{jk}v^\lambda_k\cf_{\lambda\nu},
\ i=1,\ldots,\dim\G, \label{klimseveq}
\end{equation}where $v_i$ form basis of left--invariant fields on $\G$ and
$\tilde f_i^{jk}$ are structure coefficients of th Lie group $\tG$.
Explicit form of dualizable tensors $\cf$ and their \PL transformed forms are given in the Appendix.

The NS-NS \cor{part} of \sugra s  ${GSE(\cf,\Phi,J)}$ read \cite{Wulff:2016tju,sugra2}
\begin{align}\label{betaG}
0 &= R_{\mu\nu}-\frac{1}{4}H_{\mu\rho\sigma}H_{\nu}^{\
\rho\sigma}+\nabla_{\mu}X_{\nu}+\nabla_{\nu}X_{\mu},\\ \label{betaB}
0 &=
-\frac{1}{2}\nabla^{\rho}H_{\rho\mu\nu}+X^{\rho}H_{\rho\mu\nu}+\nabla_{\mu}X_{\nu}-\nabla_{\nu}X_{\mu},\\
\label{betaPhi} 0 &=
R-\frac{1}{12}H_{\rho\sigma\tau}H^{\rho\sigma\tau}+4\nabla_{\mu}X^{\mu}-4X_{\mu}X^{\mu}
\end{align} where $R_{\mu\nu}$ is Ricci tensor of metric $\mathcal G$, $R=R_\mu{}^\mu$,
\begin{equation}
 H_{\rho\mu\nu}=\partial_\rho {\mathcal {B}}_{\mu\nu}+\partial_\mu {\mathcal {B}}_{\nu\rho}+\partial_\nu
{\mathcal {B}}_{\rho\mu}
\end{equation} and 
\begin{equation}\label{xform}
X_\mu:=\partial_\mu\Phi+J^\kappa\cf_{\kappa\mu}.
\end{equation}
\cor{For the NS-NS part of \sugra s \eqref{betaG}-\eqref{betaPhi} it is not necessary to require that $J$ be Killing vector field of $\mathcal G,H,\phi$ even though it is it is required for their full version containing the R-R fields.
Goal of this section} is to define \PL \tfn s of the fields $\phi,J$ that keep \cor{the \eqn s \eqref{betaG}-\eqref{betaPhi}} satisfied.

\cor{Let \dd\ has two decomposition }$\D=(\G|\tG)=(\wh\G|\bar\G)$. Transformation of dilaton under \pltp y can be expresed as
\begin{equation}
\Phi^0(s,y)=:\Phi(s,y)-\half L(s,y)=\wh\Phi(s,\hat x)-\half\wh L(s,\hat x)
 \label{dualdil1}
\end{equation}
where $\Phi(s,y), \wh\Phi(s,\hat x)$ are dilatons of the initial and transformed model. Variables $y$ represent \coor s of group $\G$, $\hat x$ are \coor s of group $\wh \G$,  and terms $L(s,y)$, $\wh L(s,\hat x)$ read
\begin{equation}
L(s,y) =\ln\Big| \frac{(\det \mathcal G(s,y))^{1/2}}{\det u(y)}\Big|, \quad
\wh L(s,\hat x) = \ln \Big| \frac{(\det \wh{\mathcal{G}}(s,\hat x))^{1/2}}{\det \hat u(\hat x)}\Big|,
\end{equation}
where $\mathcal{G}$ and $\wh{\mathcal{G}}$ are metrics of \sm s on  $\N\times\G$ resp. $\N\times\wh \G$. Matrices  $u$, $\hat u$ are  components of left-invariant forms of $\G$ and $\wh \G$. 

In cases when the invariant dilaton $\Phi^0$ in \eqref{dualdil1} depends on coordinates $y$ we have to express $y$ in terms of $\hat x$ and $\bar x$ to get explicit form of transformed dilaton $\wh\Phi(s,\hat x)$.
This can be done by
solving relation between two different decompositions of elements of \dd\  $\D=(\G|\tG)=(\wh\G|\bar\G)$.  
\begin{equation}
\label{lgh}
g(y)\wt h(\wt y)=\wh g(\hat x)\bar h(\bar x),\quad 
g\in \G,\ \wt h\in \wt\G,\ \wh g\in \wh\G,\ \bar h\in \bar\G
\end{equation}
so that
\begin{equation}\label{yInx}
y^k=Y^k(\hat x,\bar x),\quad \tilde y^k=\wt Y^k(\hat x,\bar x). 
\end{equation} 

If $\Phi^0(s,y)$ after the insertion  \eqref{yInx} into \eqref{dualdil1} dependends linearly on dual-\coor s $\bar x_a$ 
\begin{equation}\label{dualizability of X}
\wh\Phi^0(s,\hat x,\bar x):= \Phi^0(s,Y(\hat x,\bar x))=\wh\Phi^0(s,\hat x)+\bar d^a\,\bar x_a
\end{equation}
then we can transform dilaton $\Phi$  and vector field $J$ 
in the following way (see \cite{demhast,saka2}).
\begin{equation}
\wh\Phi(s,\hat x)=\wh \Phi^0(s,\hat x)+\half\wh L(s,\hat x)
\label{dualdil}
\end{equation} 
$$
\wh{\mathcal{J}}^\alpha=0,\qquad \alpha=1,\ldots, n=\dim \N,
$$
\begin{equation}
\label{kilJ}
  \mathcal{\wh J}^{\dim \N+ m}(s,\hat x)=
     \left(\half{{\bar f}^{ab}}{}_b -\bar d^a\right){\wh v_a}{}^m(\hat x)
\end{equation}

The above  formulas work well  for isometric models, i.e. if $\tG$ is abelian. For some more general type of models transformation of Killing field 
must be extended to \cite{hlape:formula for J}
$$
\wh{\mathcal{J}}^\alpha=0,\qquad \alpha=1,\ldots, n=\dim \N,
$$
\begin{align}
\nonumber
 \mathcal{\wh J}^{\dim \N+ m}(s,\hat x)= &
 \half{{\widetilde f}^{ab}}{}_b \left(\frac{\partial{\widetilde  y_a}}{\partial\bar x_k}{\wh v_k}{}^m(\hat x)-\frac{\partial{\widetilde  y_k}}{\partial\wh x^a} \wh\cf^{km}\right)+\\ 
   & + \left(\half{{\bar f}^{ab}}{}_b -\bar d^a\right){\wh v_a}{}^m(\hat x)
\label{kilJ lhip}\end{align}
where $a,b,k,m=  1,\ldots,\dim \G$, 
${\widetilde f}^{ba}{}_c$ and ${\bar f}^{ba}{}_c$ are
structure constants of Lie algebras of $\tG,\,\bG$ and $\wh  v_{a}$ are left-invariant fields of the 
group $\hG$. This modification does not change results of \cite{saka2,hokico} and \cite{hlape:pltpbia} because those papers deal with isotropic initial models whose corresponding Manin triples  $(\cd, \cg, \tcg)$ are semiabelian, i.e. ${{\widetilde f}^{ab}}{}_b=0$.

\section{Symmetries of \sugra s}\label{symmetries}
Let $(\cf,\Phi,J)$ satisfy \sugra s and there is a a symmetry of these \eqn s, i.e. \tfn\ $(\cf,\Phi,J)\mapsto (\cf',\Phi',J') $ that keep the \sugra s satisfied
\begin{equation}\label{sugra covariance2}
GSE(\cf,\Phi,J) \Leftrightarrow GSE(\cf',\Phi',J')
.\end{equation}
We shall call this symmetry  \emph{\PL compatible} if $(\wh\cf',\wh\Phi',\wh J')$ obtained by \PL\ \tfn\ of ($\cf',\Phi,J')$ satisfy \sugra s as well, i.e. if 
\begin{equation}\label{PL compatibility}
GSE(\cf',\Phi',J') \Leftrightarrow GSE(\wh\cf',\wh\Phi',\wh J')
.\end{equation}

Our aim is 
finding symmetries compatible with \PL \tfn s 
 \eqref{dualdil} and \eqref{kilJ} 
or \eqref{kilJ lhip}.

\subsection{$\chi$-symmetry} 
First symmetry we are going to investigate is shift of form $X$. \cor{For torsionless backgrounds, which are all examples below,} it is easy to see  that if $X_\mu$ satisfy the \sugra s, then
\begin{equation}
\label{ambig X} X'_\mu:=X_\mu +\chi_\mu,
\end{equation}
where
\begin{equation}
\label{cond for chi} \nabla_\nu\chi_\mu=0, \quad (X_\mu +2\,\chi_\mu)\chi^\mu=0,
\end{equation}
satisfy the equations as well. Due to the former \cond\ form $\chi$ is (locally) exact so that $\chi=d \psi $ and this symmetry is just ($t,x$-dependent) shift of dilaton \begin{equation}\label{dilaton shift}
\Phi'=\Phi+\psi.
\end{equation} Note that the vector field $J$ remains unchanged.

Unfortunately, in many cases $\chi$-symmetries are not \PL compatible.

\subsubsection{Example 1}\label{Example 2}
Solving the \eqn s \eqref{cond for chi} for flat metric\footnote{We work with four-dimensional models invariant w.r.t. three-dimensional groups so that $dim \N=1$, and spectator is denoted as $t$.}
\begin{equation}\label{flatB5}
ds^2=-dt^2+t^2\,dy_1^2+t^2\,\e^{2y_1}dy_2^2+t^2\,\e^{2y_1}dy_3^2
\end{equation}
with \coor s adapted to its Bianchi 5 symmetry 
\begin{equation}
[T^1,T^2]  = T^2,\quad  [T^1,T^3]  = T^3,
\end{equation}
and dilaton $\Phi=0$ we get 
\begin{equation}\label{psi51}
\chi_\mu= (C_1\, e^{y_1},C_1\, t\, e^{y_1},0,0),\quad\Phi'=\psi=C_1\, t\, e^{y_1}+C_0
\end{equation}
By nonabelian T-duality given by the matrix 
\begin{equation}\label{excmat}
C=\left(
\begin{array}{cccccc}
 0 & 0 & 0 & 1 & 0 & 0 \\
 0 & 0 & 0 & 0 & 1 & 0 \\
 0 & 0 & 0 & 0 & 0 & 1 \\
 1 & 0 & 0 & 0 & 0 & 0 \\
 0 & 1 & 0 & 0 & 0 & 0 \\
 0 & 0 & 1 & 0 & 0 & 0 \\
\end{array}
\right)
\end{equation}  we get (see Appendix) 
\begin{equation}\label{dualflat}
\wh\cf_{\mu\nu}=\left(
\begin{array}{cccc} -1 & 0 & 0 & 0 \\
 0 & \frac{t^2}{t^4+{\hat x}_2^2+{\hat x}_3^2} & \frac{{\hat x}_2}{t^4+{\hat x}_2^2+{\hat x}_3^2} & \frac{{\hat x}_3}{t^4+{\hat x}_2^2+{\hat x}_3^2} \\
 0 & -\frac{{\hat x}_2}{t^4+{\hat x}_2^2+{\hat x}_3^2} & \frac{t^4+{\hat x}_3^2}{t^2 \left(t^4+{\hat x}_2^2+{\hat x}_3^2\right)} & -\frac{{\hat x}_2 {\hat x}_3}{t^2
   \left(t^4+{\hat x}_2^2+{\hat x}_3^2\right)} \\
 0 & -\frac{{\hat x}_3}{t^4+{\hat x}_2^2+{\hat x}_3^2} & -\frac{{\hat x}_2 {\hat x}_3}{t^2 \left(t^4+{\hat x}_2^2+{\hat x}_3^2\right)} & \frac{t^4+{\hat x}_2^2}{t^2
   \left(t^4+{\hat x}_2^2+{\hat x}_3^2\right)} \\
\end{array}
\right) 
\end{equation}
but if $C_1\neq  0$ \emph{we cannot apply formulas \eqref{dualdil} and \eqref{kilJ} or \eqref{kilJ lhip} for the non-Abelian T-duality} (and some other \pltp ies) as  the \cond\ \eqref{dualizability of X} for its application does not hold 
because
$$ \wh\Phi^0\,'(t,\hat x,\bar x)=
C_1\, t\, e^{\bar x_1}+C_0 +\bar x_1 -\frac{3}{2} \log\,t. $$ 

\subsubsection{Example 2}\label{Example 3}
Background 
 \begin{equation}\label{flat4}
\cf_{\mu\nu}=\left(
\begin{array}{cccc}
 1 & 0 & 0 & 0 \\
 0 & 0 & e^{-y_1} y_1 & e^{-y_1} \\
 0 & e^{-y_1} y_1 & e^{-2 y_1} & 0 \\
 0 & e^{-y_1} & 0 & 0 \\
\end{array}
\right)
\end{equation}
that corresponds to the flat metric  adapted to Bianchi 4 symmetry 
satisfy \sugra s together with vanishing X-form.
Solving the \eqn s \eqref{cond for chi} we get 
\begin{align}\nonumber
\Phi'=\psi=&C_0+{C_1}\, t+{C_2} e^{-y_1}+{C_3} \left(y_1+e^{-y_1}
   y_2\right)\\&+{C_4} \left(\frac{1}{2} e^{-y_1} y_2^2+\left(y_1-1\right)
   y_2-\frac{e^{y_1}}{2}+y_3\right)\label{dualB4}
\end{align}
where the latter condition of \eqref{cond for chi} implies
$$C_1{}^2+C_3{}^2-2\,C_2C_4=0. $$
The \bkg\ \eqref{flat4}, dilaton \eqref{dualB4} and vanishing $J$ satisfy \sugra s but once again 
  we cannot apply formulas \eqref{dualdil} and \eqref{kilJ} or \eqref{kilJ lhip} for non-Abelian T-duality 
 as  the \cond\ \eqref{dualizability of X} is satisfied 
only if $\Phi'=C_0.$

\subsection{Gauge \tfn s}
Another symmetry of NS-NS \sugra s is gauge \tfn\footnote{{Note that this symmetry is different from the the \tfn\ of B-field $B_\lambda= B+d\lambda$ investigated e.g. in \cite{jabbari 1708.,jabbari 1710.}  }}
\begin{equation}\label{gauge tfn lambda}
\cf_\Lambda:=\cf,\quad\Phi_\Lambda:=\Phi +\Lambda\quad J_\Lambda:= J-d\Lambda.\cf\-1
\end{equation}
where $\Lambda$ is arbitrary (differentiable) function of $(s,y)$. It leaves $X$ invariant, $X=X_\Lambda$, so that 
\begin{equation}\label{gauge invariance 0}
 {GSE(\cf_\Lambda,\Phi_\Lambda,J_\Lambda)}\Leftrightarrow GSE(\cf,\Phi,J).
\end{equation}
Contrary to this,  
formulas \eqref{dualdil} and \eqref{kilJ} or \eqref{kilJ lhip} for \PL \tfn\ of dilaton and Killing field \emph{do not provide solution of \sugra s for arbitrary $\Lambda$.}
$$ GSE(\cf_\Lambda,\Phi_\Lambda,J_\Lambda) <\neq >{GSE(\wh\cf_\Lambda,\wh \Phi_\Lambda,\wh J_\Lambda)}.$$

\subsubsection{Example 3, trivial - identical \tfn\ of flat \bkg}
\label{trivial ex}
Let us investigate the simplest \PL\ \tfn\ - identity  of the flat model \eqref{flatB5} with $\mathcal B$-field vanishing. 
This \bkg\ together with \begin{equation}\label{XB5}
\Phi=0,\quad J^\mu=(0,0,0,0), \quad X_\mu=(0,0,0,0),
\end{equation} obviously satisfy \sugra s. Besides that, identical \PL \tfn\ of \eqref{XB5} gives the same fields $\phi,J,X$.

However, \sugra s  are satisfied also  for 
\begin{equation}\label{fijlam}
\Phi_\Lambda= \Lambda, \quad J^\mu_\Lambda=-\partial_\nu\Lambda\,\G^{\nu\mu},\quad \cf_\Lambda=\cf, \quad  X_\Lambda=(0,0,0,0)
\end{equation}
where $\Lambda$ is arbitrary  function of $(t,y_1,y_2,y_3)$. Choosing for example $\Lambda=y_1$ we get
\begin{equation}\label{fijlam1}\Phi=y_1,\ J^\mu=(0,\frac{-1}{t^2},0,0).
\end{equation}
Applying formulas \eqref{dualdil} and \eqref{kilJ} for identical \PL \tfn\  to \eqref{fijlam1} we find
$$\wh\Phi=\hat x_1,\quad\wh J^\mu= (0,0,0,0),\quad \wh\cf_\Lambda=\cf, \quad  \wh X_\Lambda=(0,1,0,0) $$
and \sugra s are not satisfied. 

This simple example shows that \emph{\PL\ \tfn s are not in general compatible with 
gauge \tfn s} and 
that the \cond\ \eqref{dualizability of X} is not sufficient for applicability of the formulas \eqref{dualdil} and \eqref{kilJ} or \eqref{kilJ lhip}. 

Note that the field $J$ in \eqref{fijlam1} in the latter case is not Killing field of the flat metric.
It may give a clue for restriction of the gauge \tfn s.

\section{Choice of \PL\ compatible gauge}
We have seen that in spite of the fact that due to symmetries there can be quite large set of dilatons and vector fields $J$ that 
satisfy \sugra s, only very limited subset of them 
are \PL compatible. 
By inspection of the formulas \eqref{dualdil} and \eqref{kilJ} one can see that  the problem is in fulfilling the \cond\ \eqref{dualizability of X} for invariant dilaton $\Phi^0$ shifted both by $\chi$-symmetry and gauge \tfn
\begin{equation}\label{dualdil0psilam}
\Phi^{0}\,'(s,y)=\Phi^0(s,y)+\psi(s,y)+\Lambda(s,y).
\end{equation} 
Fortunately, we can use the arbitrariness of the gauge function $\Lambda$ to satisfy the \cond\ \eqref{dualizability of X}. On the other hand, we know from the example \ref{trivial ex} that gauge \tfn s in general are not compatible with \PL\ \tfn\ 
so that they must be further restricted.

Condition that the field $J$ is  Killing vector  of the \bkg\ $\cf= {\mathcal {G}}+{\mathcal {B}}$ (up to an exact 2-form)  is not necessary for satisfying NS-NS \sugra s but it is required for their full version containing the R-R fields \cite{sugra2}. Beside that, dilaton must also be invariant  in direction of $J$. Therefore, if $(\cf,\Phi,J)$ satisfy \sugra s \eqref{betaG}-\eqref{betaPhi}  we will require for $\Lambda$ 
\begin{equation} \label{kilingovostF}
\mathcal L_{J_\Lambda}{\mathcal {G}}=0,\quad \mathcal L_{J_\Lambda}{\mathcal {B}}=d\omega,\quad \mathcal L_{J_\Lambda} \Phi=0
\end{equation}
where and $J_\Lambda:= J-d\Lambda.\cf\-1$. It turns out that \emph{these additional \cond s together with \eqref{dualizability of X} are sufficient for compatibility of \PL \tfn s with symmetries introduced in the Sec. \ref{symmetries}.}
\subsection{Example 1 - continued}\label{Example 2 cont}
Let $\mathcal {B} $-field is vanishing, \bkg\ is given  by the flat metric \eqref{flatB5}, vanishing $J$-field  and dilaton obtained by $\chi$-symmetry and gauge \tfn\ 
\begin{equation}\label{dilchi23}
\Phi'(t,y)=C_1\, t\, e^{y_1}+C_0 +\Lambda(t,y_1,y_2,y_3).
\end{equation} They satisfy \sugra s. Condition \eqref{dualizability of X} for non-Abelian T-duality is fulfilled for 
\begin{equation}
\Lambda=-C_1\, t\, e^{y_1} + \Lambda_0(t)+\lambda_1y_1+\lambda_2\, y_2+\lambda_3\, y_3.
\end{equation}
Requiring that $J_\Lambda$ is Killing vector of flat metric \eqref{flatB5},  and dilaton \eqref{dilchi23} 
we get 
$$\Lambda_0(t)=\lambda_1=\lambda_2=\lambda_3=0.$$ This gauge is  compatible with T-duality of \bkg\ \eqref{flatB5} dilaton \eqref{dilchi23} and vanishing $J$-field. This means that only the trivial dilaton $\Phi'=C_0$ can be dualized by formulas \eqref{dualdil}, \eqref{kilJ}.  
By the non-Abelian T-dual given by \eqref{excmat} we get the \bkg\ \eqref{dualflat} and formulas  \eqref{dualdil}, \eqref{kilJ} yield \cite{hlape:pltdid}
\begin{equation}\label{dilJex2}
\wh\Phi=-\frac{1}{2} \log \left(-t^2 \left(t^4+{\hat x}_2^2+{\hat x}_3^2\right)\right),\quad \wh J^\mu=(0,2,0,0).
\end{equation}
\sugra s are satisfied for these fields.
 
Let us note that repeating the \PL \tfn\  given by \eqref{excmat} on the tensor field \eqref{dualflat} we return to flat metric \eqref{flatB5} but dual dilaton and vector $J$ given by\eqref{dualdil} and \eqref{kilJ lhip} are of the form \eqref{fijlam1} that differ from the initial ones \eqref{XB5} by gauge \tfn\ $\Lambda=y_1$.

\subsection{Example 2 - continued}\label{Example 3 cont}
Let $\mathcal {B} $-field is vanishing, \bkg\ is given  by the flat metric \eqref{flat4}, vanishing $J$-field  and by dilaton 
\begin{align}\label{dualdilB4}
\Phi'=&C_0+{C_1}\, t+{C_2} e^{-y_1}+{C_3} \left(y_1+e^{-y_1}
   y_2\right)\\&+{C_4} \left(\frac{1}{2} e^{-y_1} y_2^2+\left(y_1-1\right)
   y_2-\frac{e^{y_1}}{2}+y_3\right)+\Lambda(t,y_1,y_2,y_3).
\end{align}
They satisfy \sugra s. Condition \eqref{dualizability of X} for T-duality given by \eqref{excmat} is fulfilled if 
\begin{align}
\Lambda=&{c_0}+{c_1} y_2+{c_2} y_3+\frac{1}{2} \left({C_4}
   e^{y_1}-e^{-y_1} \left(2 {C_2}+2 {C_3} y_2+{C_4}
   y_2^2\right)\right)+\\ &+y_1 \left({c_4}-{C_4}
   y_2\right)+\Lambda_0(t).
\end{align}
Requiring that $J_\Lambda$ is Killing vector of flat metric \eqref{flat4},  and dilaton \eqref{dualdilB4} 
one gets 
$$c_1=- c_2=C_4, \quad c_4=-C_3, \quad\Lambda_0(t)={c_5}-{C_1} t.$$
This gauge eliminates  $\chi$-symmetry shift in  \eqref{dualdilB4} up to constant and only the trivial dilaton $\Phi'=C_0$ can be dualized by formulas \eqref{dualdil}, \eqref{kilJ}. We get 
\begin{equation}
\wh\Phi=\wh C_0-\half \log \left(\hat x_3{}^2-1\right),\quad \wh J=(0,-2,0,0)
\end{equation} that together with
\begin{equation}
\wh\cf=\left(
\begin{array}{cccc}
 1 & 0 & 0 & 0 \\
 0 & 0 & 0 & \frac{1}{1-\hat x_3} \\
 0 & 0 & 1 & \frac{\hat x_3-\hat x_2}{\hat x_3-1} \\
 0 & \frac{1}{\hat x_3+1} & \frac{\hat x_3-\hat x_2}{\hat x_3+1} &
   \frac{(\hat x_2-\hat x_3)^2}{\hat x_3{}^2-1} \\
\end{array}
\right),
\end{equation}obtained by \eqref{Fhat}-\eqref{Pihat}, satisfy \sugra s.

On the other hand, let us note that there are \PL \tfn s that impose weaker restriction on the gauge \tfn s and therefore admit a wider subset of dilatons that can be \PL\ transformed. It is for example
\pltp y $(4|1)\rightarrow(6_{-1}|2)$ of \eqref{flat4}, \eqref{dualdilB4} given by
\begin{equation}
 C=\left(
\begin{array}{cccccc}
 -1 & 0 & 0 & 0 & 0 & 0 \\
 0 & 0 & 0 & 0 & 1 & 0 \\
 0 & 0 & -1 & 0 & 0 & 0 \\
 0 & 0 & 0 & -1 & 0 & 0 \\
 0 & 1 & 0 & 0 & 0 & 0 \\
 0 & 0 & 0 & 0 & 0 & -1 \\
\end{array} \right)
\label{41to612}
\end{equation}
Condition \eqref{dualizability of X} then gives $C_1=C_3=C_4=0$ and 
\begin{equation}
\Lambda
=\Lambda_1(t,y_1,y_3)+\Lambda_2(y_1,y_3)+\Lambda_3(y_1)+ c_1y_2.
\end{equation}
Requiring further that $J_\Lambda$ is Killing vector of flat metric \eqref{flat4},  and dilaton \eqref{dualdilB4}, i.e. satisfies \eqref{kilingovostF}, 
one gets finally
\begin{equation}\label{dil4}
\Lambda=c_2+c_3 e^{-y_1},\quad \Phi'=C_0+{C_2}\, e^{-y_1}+c_2+c_3 e^{-y_1}
\end{equation}
All these dilatons can be pluralized.
\pltp y induced by \eqref{41to612} then gives
\begin{equation}
\wh\cf=\left(
\begin{array}{cccc}
 1 & 0 & 0 & 0 \\
 0 & -{\hat x}_1^2 & -e^{-x_1} {\hat x}_1 & e^{{\hat x}_1} \\
 0 & e^{-{\hat x}_1} {\hat x}_1 & e^{-2 {\hat x}_1} & 0 \\
 0 & e^{{\hat x}_1} & 0 & 0 \\
\end{array}
\right)
\end{equation}
\begin{equation}
\wh\Phi=C_0+(C_2+c_3)\, e^{{\hat x}_1},\quad \wh J=(0,0,0,0)
\end{equation} and \sugra s are satisfied.

\section{Conclusion}
We have investigated two types of \tfn s that keep  NS-NS part of \sugra s satisfied. They are $\chi$-symmetry \eqref{ambig X} \cor{for torsionless \sm s}, that shifts dilaton only, and gauge \tfn s \eqref{gauge tfn lambda} that change both dilaton and vector field $J$ but leave the form 
$$ X_\mu:=\partial_\mu\Phi+J^\kappa\cf_{\kappa\mu}.
 $$
invariant. Due to these symmetries there is a large set of dilatons and vector fields $J$ that satisfy \sugra s for fixed tensor field $\cf$. 

We have shown that \PL \tfn s 
are not in general compatible with the above mentioned symmetries - see Examples 1,2,3. In other words, formulas \eqref{dualdil}, \eqref{kilJ} or \eqref{kilJ lhip} for \PL \tfn s of dilatons and vector fields $J$ can be applied only to a rather narrow subset of dilatons in order that the transformed fields  satisfy \sugra s.  

The applicability of the formulas 
\eqref{dualdil}, \eqref{kilJ} or \eqref{kilJ lhip} 
is restricted \\ 
1) By the \cond\    \eqref{dualizability of X} requiring that the invariant dilaton $\Phi^0$ given by \eqref{dualdil1} is linear in the dual \coor s $\bar x$.\\
2) By the \cond s \eqref{kilingovostF} that fixes the admissible gauges, namely, that  gauge transformed vector field $J_\Lambda$ is Killing vector \cor{of metric, torsion and dilaton. It is interesting that this condition of \PL compatibility is identical with condition for full \sugra s containing the R-R fields}.

Within these restrictions \PL \tfn s  keep \sugra s satisfied - see Examples 1,2 continued. Typically it chooses just one dilaton and Killing vector field but it is not a rule as shown in the subsection \ref{Example 3 cont}. We have checked several other cases of \PL duality/plurality with equal results.

Besides that we have found that twice applied T-duality produces identical dilatons but vector fields $J$ 
only up to a gauge \tfn. It means that, diferently from \tfn s of tensor fields \eqref{Fhat} - \eqref{Pihat}, the above mentioned formulas do not provide true representation of $O(d,d)$.

\section{Appendix: \PL \tfn s of the tensor field}
For many \dd s several decompositions may exist.
Suppose that we have \sm\ on $\N\times \G$ and tensor field $\cf$ satisfies \eqn\ (\ref{klimseveq}).   Let \dd {}   $\D=(\G|\tG)$ splits
into another pair of Lie subgroups $\hG$ and $\bG$ so that $(\G|\tG)=(\wh\G|\bar\G)$. Then we can apply
the full framework of \PL\ T-plurality \cite{klise, unge:pltp} and
find tensor field $\wh\cf$ for sigma model on $\N \times \hG$ in the following way.

\PL\ dualizable sigma models on $\N \times \G$ satisfying (\ref{klimseveq}) are given by tensor field $\cf$ of the form\begin{equation}\label{F}
\cf(s,y)=\mathcal{E}(y)\cdot\left(\unit_{n+d}+E(s) \cdot \Pi(y)\right)^{-1}\cdot
E(s)\cdot \mathcal{E}^T(y)
\end{equation}
where $E(s)$ is spectator-dependent $(n+d)\times (n+d)$ matrix. Denoting generators of Manin triple $(\cd, \cg, \tcg)$ as $T, \wt T$, matrix $\Pi(y)$ is given by submatrices $a(y)$ and $b(y)$ of the adjoint representation
$$
ad_{{g}\-1}(\widetilde T) = b(y) \cdot T + a^{-1}(y)\cdot \widetilde T
$$
as
$$\Pi(y)= \left(
\begin{array}{cc}
\nul_n & 0 \\
 0 &b(y) \cdot a^{-1}( y)
\end{array}
\right).$$
Matrix $\mathcal{E}(y)$ reads 
\begin{equation}\label{eRextended}
\mathcal{E}(y)=
\left(
\begin{array}{cc}
 \unit_n & 0 \\
 0 & e(y)
\end{array}
\right)
\end{equation}
where $e(y)$ is $d\times d$ matrix of components of right-invariant Maurer--Cartan form $(dg)g^{-1}$  on $\G$.

Manin triples $(\cd, \cg, \tcg)$ and
$(\cd,\hcg, \bcg)$ are two decompositions  of Lie algebra $\cd$ into double cross sum 
of subalgebras  that are maximally isotropic with respect to
$\langle . , . \rangle$.
Pairs of mutually dual bases $T_a \in \cg,\
\widetilde{T}^a \in \tcg$ and $\wh T_a \in \hcg,\ \wb{T}^a \in
{\bcg}$, $a=1, \ldots, d,$ 
then must be related by transformation
\begin{equation}\label{C_mat}
\begin{pmatrix}
\wh T \\
\wb T
\end{pmatrix}
 = C \cdot
\begin{pmatrix}
T \\
\widetilde T
\end{pmatrix}
\end{equation}
where $C$ is an invertible $2d\times 2d$ matrix. (Non-Abelian) T-duality is obtained by 
$$C=\begin{pmatrix}
 0 & \unit_d \\
 \unit_d & 0
\end{pmatrix}.$$
\pltp y is given by $d \times d$ matrices $P, Q, R, S$ such that
\begin{equation}\label{pqrs}
\begin{pmatrix}
T \\
\widetilde T
\end{pmatrix}
= C^{-1} \cdot
\begin{pmatrix}
\wh T \\
\wb T
\end{pmatrix} =
\begin{pmatrix}
 P & Q \\
 R & S
\end{pmatrix} \cdot
\begin{pmatrix}
\wh T \\
\wb T
\end{pmatrix}.
\end{equation}
For the following formulas it is convenient to extend  matrices $P, Q, R, S$ to $(n+d)\times (n+d)$ matrices
\begin{equation}
\label{pqrs2}
\nonumber
\mathcal{P} =\begin{pmatrix}\unit_n &0 \\ 0&P \end{pmatrix}, \quad \mathcal{Q} =\begin{pmatrix}\nul_n&0 \\ 0&Q \end{pmatrix}, \quad \mathcal{R} =\begin{pmatrix}\nul_n&0 \\ 0&R \end{pmatrix}, \quad \mathcal{S} =\begin{pmatrix}\unit_n &0 \\ 0& S \end{pmatrix}
\end{equation}
{to accommodate the spectator fields.} 

Sigma model on $\N \times \hG$ obtained from \eqref{F} via \pltp y
is given by tensor field
\begin{equation} \label{Fhat} \widehat{\cf}(s,\hat x)=
\mathcal{\widehat E}(\hat x)\cdot \widehat E(s,\hat x) \cdot
\mathcal{\widehat E}^T(\hat x), \qquad \mathcal{\widehat E}(\hat x)=
\begin{pmatrix}
\unit_n & 0 \\
 0 & \wh e(\hat x)
\end{pmatrix},
\end{equation}
where $\wh e(\hat x)$ is $d\times d$ matrix of components of
right-invariant Maurer--Cartan form $(d\hat g)\hat g^{-1}$  on
$\wh\G$ and
\begin{equation}\label{Fhat2} \wh
E(s,\hat x)=\left(\unit_{n+d}+\wh E(s) \cdot \wh{\Pi}(\hat x)\right)^{-1}\cdot
\wh E(s)  =\left(\wh E\-1(s)+ \wh{\Pi}(\hat x)\right)^{-1}.
\end{equation}
The matrix $\wh E(s)$ is obtained from $E(s)$ in \eqref{F} by formula
\begin{equation}\label{E0hat}
\wh E(s)=(\mathcal{P}+ E(s) \cdot \mathcal{R})^{-1} \cdot
(\mathcal{Q}+E(s) \cdot \mathcal{S}),
\end{equation}
and
$$\widehat\Pi(\hat x)= \left(
\begin{array}{cc}
\nul_n & 0 \\
 0 & \widehat b(\hat x) \cdot \widehat a^{-1}(\hat x)
\end{array}
\right),$$
\begin{equation}\label{Pihat}
ad_{{\hat g}\-1}(\wb T) = \widehat b(\hat x) \cdot \wh T + \wh a^{-1}(\hat x)\cdot \wb T.
\end{equation}

\end{document}